\documentclass{sigchi}

\CopyrightYear{2017} 
\setcopyright{acmcopyright} 
\conferenceinfo{UIST 2017,}{October 22--25, 2017, Quebec City, QC, Canada}
\isbn{978-1-4503-4981-9/17/10}\acmPrice{\$15.00}
\doi{https://doi.org/10.1145/3126594.3126653}

\usepackage{balance}  
\usepackage{graphics} 
\usepackage[T1]{fontenc}
\usepackage{txfonts}
\usepackage{mathptmx}
\usepackage[pdflang={en-US},pdftex]{hyperref}
\usepackage{color}
\usepackage{tabu}
\usepackage{booktabs}
\usepackage{textcomp}
\usepackage{arydshln}

\usepackage{microtype} 
\usepackage{ccicons}  

\usepackage{float}

\newcommand{\rev}[1]{{\color{black}{#1}}}

\makeatletter
\def\blfootnote{\xdef\@thefnmark{}\@footnotetext}
\makeatother

\usepackage{todonotes}

\def\plaintitle{Learning Visual Importance  \\ for Graphic Designs and Data Visualizations}
\def\plainauthor{First Author, Second Author, Third Author,
  Fourth Author, Fifth Author, Sixth Author}

\def\plainkeywords{Saliency; Computer Vision; Machine Learning; Eye Tracking; Visualization; Graphic Design; Deep Learning; Retargeting.}

\makeatletter
\def\url@leostyle{%
  \@ifundefined{selectfont}{
    \def\UrlFont{\sf}
  }{
    \def\UrlFont{\small\bf\ttfamily}
  }}
\makeatother
\def\pprw{8.5in}
\def\pprh{11in}

\definecolor{linkColor}{RGB}{6,125,233}
\hypersetup{%
  pdftitle={\plaintitle},
  pdfauthor={\plainauthor},
  pdfkeywords={\plainkeywords},
  pdfdisplaydoctitle=true, 
  bookmarksnumbered,
  pdfstartview={FitH},
  colorlinks,
  citecolor=black,
  filecolor=black,
  linkcolor=black,
  urlcolor=linkColor,
  breaklinks=true,
  hypertexnames=false
}

\begin{document}

\title{\plaintitle}

\numberofauthors{1}
\newcommand\Mark[1]{\textsuperscript#1}
\author{%
  \alignauthor Zoya Bylinskii\Mark{1} \space\space
  Nam Wook Kim\Mark{2} \space\space
  Peter O'Donovan\Mark{3} \space\space
  Sami Alsheikh\Mark{1} \space\space
  Spandan Madan\Mark{2} \space\space
  \\
  Hanspeter Pfister\Mark{2} \space\space
  Fredo Durand\Mark{1} \space\space
  Bryan Russell\Mark{4} \space\space
  Aaron Hertzmann\Mark{4}
  \\
  \affaddr{\Mark{1} MIT CSAIL, Cambridge, MA USA}\space\email{\texttt{\{zoya,alsheikh,fredo\}}@mit.edu} \\
  \affaddr{\Mark{2} Harvard SEAS, Cambridge, MA USA}\space\email{\texttt{\{namwkim,spandan\_madan,pfister\}}@seas.harvard.edu} \\
  \affaddr{\Mark{3} Adobe Systems, Seattle, WA USA}\space\email{\texttt{\{podonova\}}@adobe.com} \\
  \affaddr{\Mark{4} Adobe Research, San Francisco, CA USA}\space\email{\texttt{\{hertzman,brussell\}}@adobe.com} \\
}


\maketitle

\begin{abstract}
Knowing where people look and click on visual designs can provide clues about how the designs are perceived, and where the most important or relevant content lies. The most important content of a visual design can be used for effective summarization or to facilitate retrieval from a database. We present automated models that predict the relative importance of different elements in data visualizations and graphic designs. Our models are neural networks trained on human clicks and importance annotations on hundreds of designs. We collected a new dataset of crowdsourced importance, and analyzed the predictions of our models with respect to ground truth importance and human eye movements. We demonstrate how such predictions of importance can be used for automatic design retargeting and thumbnailing. User studies with hundreds of MTurk participants validate that, with limited post-processing, our importance-driven applications are on par with, or outperform, current state-of-the-art methods, including natural image saliency. We also provide a demonstration of how our importance predictions can be built into interactive design tools to offer immediate feedback during the design process. 
\end{abstract}

\category{H.5.1}{Information Interfaces and Presentation}{Multimedia Information Systems} 

\keywords{\plainkeywords}

\section{Introduction}


\begin{figure}[h]
\centering
\includegraphics[width=1\linewidth]{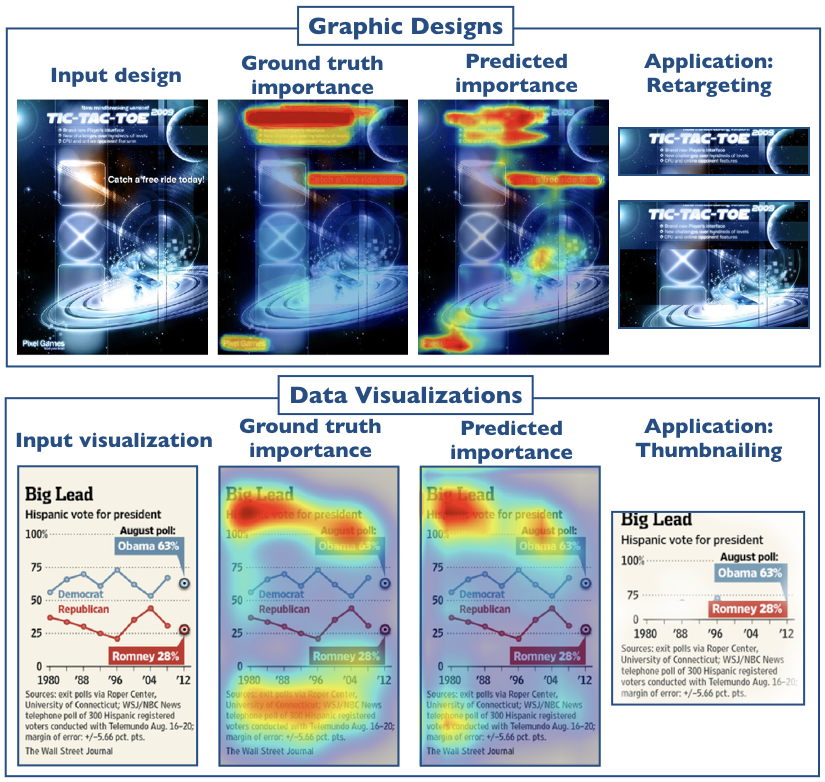}
\caption{We present two neural network models trained on crowdsourced importance. We trained the graphic design model using a dataset of 1K graphic designs with GDI annotations~\protect\cite{odonovan2014}. For training the data visualization model, we collected mouse clicks using the BubbleView methodology~\protect\cite{bubbleview} on 1.4K MASSVIS data visualizations~\protect\cite{borkin2013mem}. Both networks successfully predict ground truth importance and can be used for applications such as retargeting, thumbnailing, and interactive design tools. Warmer colors in our heatmaps indicate higher importance.
\label{fig:headerfig}
}
\end{figure}
%

A crucial goal of any graphic design or data visualization is to communicate the relative \textit{importance} of different design elements, so that the viewer knows where to focus attention and how to interpret the design. \rev{In other words, the design should provide an effective management of attention~\cite{rensink2011management}}. Understanding how viewers perceive a design could be useful for many stages of the design process; for instance, to provide feedback \cite{RosenholtzTAP}. Automatic understanding can help build tools to search, retarget, and summarize information in designs and visualizations. Though saliency prediction in natural images has recently become quite effective, there is little work in importance prediction for either graphic designs or data visualizations. 
\blfootnote{Our online demo, video, code, data, trained models, and supplemental material are available at \url{visimportance.csail.mit.edu}.}

\begin{figure*}[th]
\centering
\includegraphics[width=1\linewidth]{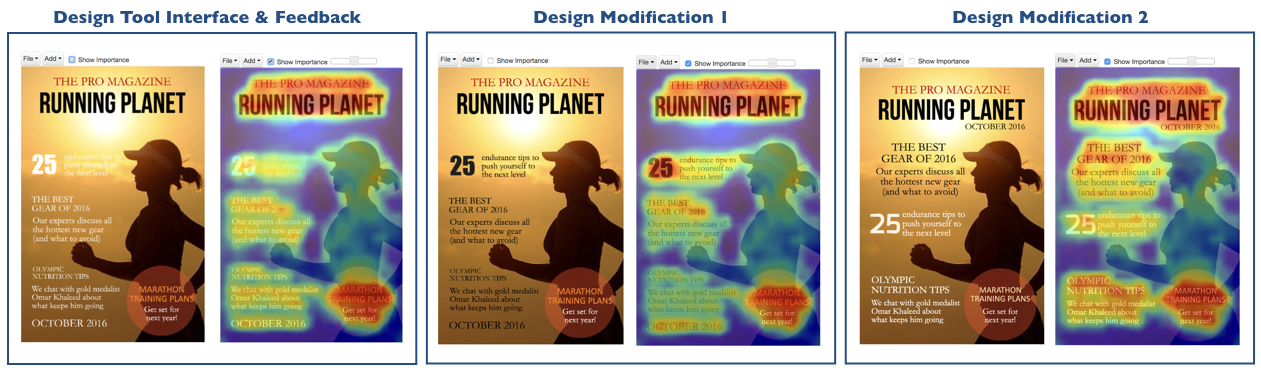}
\caption{We show an interactive graphic design application using our model that lets users change and visualize the importance values of elements. Users can move and resize elements, as well as change color, font, and opacity, and see the updated realtime importance predictions. For instance, a user changes the color of the text to the left of the runner to increase its importance (middle panel). The rightmost panel includes a few additional changes to the size, font, and placement of the text elements to modify their relative importance scores. A demo is available at \url{visimportance.csail.mit.edu}.}
\label{fig:interactiveapp}
\end{figure*}
%

We use ``importance'' as a generic term to describe the perceived relative weighting of design elements.  Image saliency, which has been studied extensively, is a form of importance. 
However, whereas traditional notions of saliency refer to bottom-up, pop-out effects,  our notion of importance can also depend on  higher-level factors such as
the semantic categories of design elements (e.g., title text, axis text, data points). 

This paper presents a new importance prediction method for graphic designs and data visualizations. We use a state-of-the-art deep learning architecture, and train models on two types of crowdsourced importance data: graphic design importance (GDI) annotations~\cite{odonovan2014} and a dataset of BubbleView clicks~\cite{bubbleview} we collected on data visualizations. 

Our importance models take input designs in bitmap form. The original vector data is not required.
As a result, the models are agnostic to the encoding format of the image and can be applied to existing libraries of bitmap designs.  
Our models pick up on some of the higher-level trends in ground truth human annotations. For instance, across a diverse collection of visualizations and designs, our models learn to localize the titles and correctly weight the relative importance of different design elements (Fig.~\ref{fig:headerfig}). 

We show how the predicted importance maps can be used as a common building block for a number of different applications, including retargeting and thumbnailing. Our predictions become inputs to cropping and seam carving with almost no additional post-processing. Despite the simplicity of the approach, our retargeting and thumbnailing results are on par with, or outperform, related methods, as validated by a set of user studies launched on Amazon's Mechanical Turk \rev{(MTurk)}.
Moreover, an advantage of the fast test-time performance of neural networks makes it feasible for our predictions to be integrated into interactive design tools (Fig.~\ref{fig:interactiveapp}). With another set of user studies, we validate that our model generalizes to fine-grained design variations and correctly predicts how importance is affected by changes in element size and location on a design.

\rev{ \textbf{Contributions:} We present two neural network models for predicting importance: in graphic designs and data visualizations. This is the first time importance prediction is introduced for data visualizations. For this purpose, we collected a dataset of BubbleView clicks on 1,411 data visualizations. 
 We also show that BubbleView clicks are related to explicit importance annotations~\cite{odonovan2014} on graphic designs. We collected importance annotations for 264 graphic designs with fine-grained variations in the spatial arrangement and sizes of design elements.
We demonstrate how our importance predictions can be used for retargeting and thumbnailing, and include user studies to validate result quality. Finally, we provide a working interactive demo.} 

\section{Related Work}

Designers and researchers have long studied eye movements as a clue to understanding the perception of interfaces~\cite{duchowski2007eye,jacob2003eye}.  There have also been several recent studies of eye movements and the perception of designs~\cite{beyondMemorability,infoAesthetics}. 
However, measuring eye movements is an expensive and time-consuming process, and is rarely feasible for practical applications. 

\begin{figure*} 
\centering
\includegraphics[width=1\linewidth]{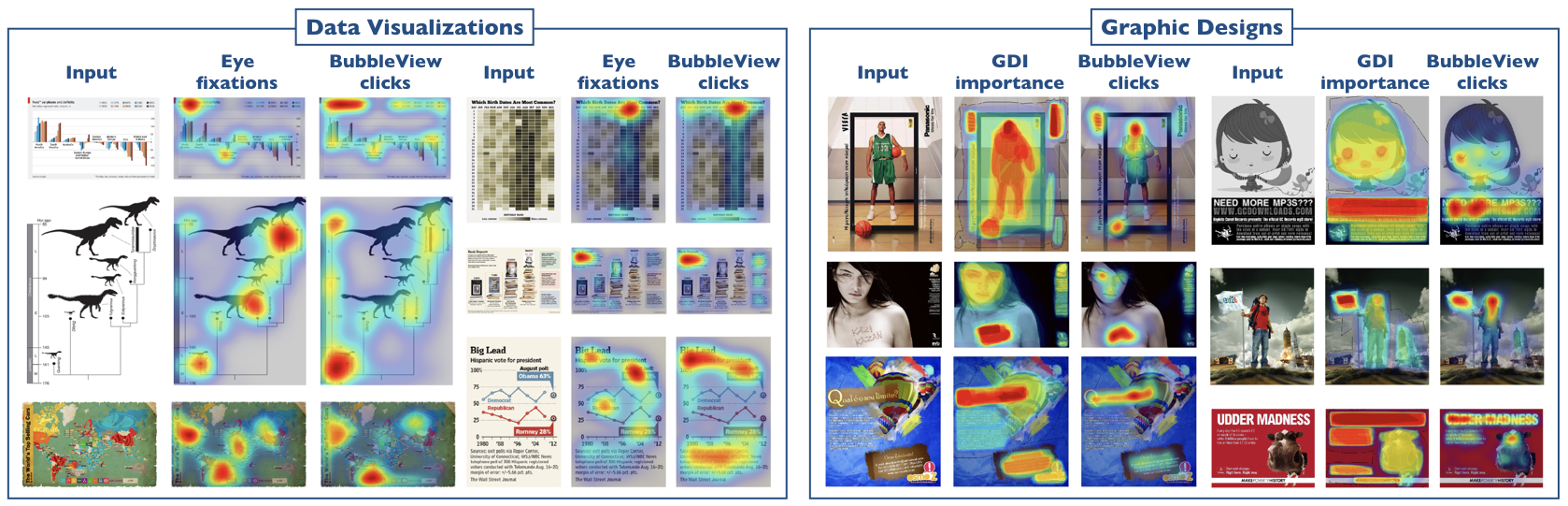}
\caption{Left: Comparison of eye movements collected in a controlled lab setting \protect\cite{beyondMemorability}, and clicks that we crowdsourced using the BubbleView interface \protect\cite{bubbleview,bubble}. 
Right: Comparison of importance annotations from the GDI dataset, and clicks that we crowdsourced using the BubbleView \rev{interface}. These examples were chosen to demonstrate some of the similarities and differences between the modalities. For instance, compared to eye fixations, clicks are sometimes more concentrated around text. Compared to the GDI annotations, clicks do not assign uniform importance to whole design elements. Despite these differences, BubbleView data leads to similar importance rankings of visual elements (\emph{Evaluation}).}
\label{fig:methodology_comparison}
\end{figure*}

Few researchers have attempted to automatically predict importance in graphic designs. The DesignEye system~\cite{RosenholtzTAP} uses hand-crafted saliency methods, demonstrating that saliency methods can provide valuable feedback in the context of a design application. O'Donovan et al.~\cite{odonovan2014} gather crowdsourced importance annotations, where participants are asked to mask out the most important design regions. They train a predictor from these annotations. However, their method requires knowledge of the location of design elements 
to run on a new design. Haass et al.~\cite{divis2016modeling} test three natural image saliency models on the MASSVIS data visualizations~\cite{borkin2013mem}, concluding that, across most saliency metrics, these models perform significantly worse on visualizations than on natural images. 
Several models also exist for web page saliency. However, most methods use programmatic elements (e.g., the DOM) as input to saliency estimation rather than allowing bitmap images as input~\cite{buscher2009you,Woodruff2001}. Pang et al. predict the order in which people will look at components on a webpage~\cite{pang2016directing} by making use of the DOM and manual segmentations. Other works use the web page image itself as input to predict saliency~\cite{shen2014webpage,still2010saliency}. 
Most of these methods use older saliency architectures based on hand-crafted features that are inferior to the state-of-the-art neural networks we use in our approach. Our work also relates to the general program of applying computer vision and machine learning in the service of graphic design tools~\cite{Kumar2,Kumar1,2011-revision}.

Predicting eye movements for natural images is a classic topic in human and computer vision. The earliest natural image saliency methods relied on hand-coded features (e.g., \cite{itti98model}). Recently, deep learning methods, trained on large datasets, have produced a substantial jump in  performance on standard saliency benchmarks~\cite{zoyaECCV2016,cornia2016deep,huang2015salicon,kummerer2014deep,pan2016shallow,zhao2015saliency}. However, these methods have been developed exclusively for analyzing natural images, and are not trained or tested on graphic designs. Our work is the first to apply neural network importance predictors to both graphic designs and data visualizations.





\section{Data Collection}

To train our models we collected BubbleView data~\cite{bubbleview,bubble} for data visualizations, and used the Graphic Design Importance (GDI) dataset by O'Donovan et al.~\cite{odonovan2014} for graphic designs. We compared different measurements of importance: BubbleView clicks to eye movements on data visualizations, and BubbleView clicks to GDI annotations on graphic designs. 




\subsection{Ground truth importance for data visualizations}

Large datasets are one of the prerequisites to train neural network models. Unfortunately, collecting human eye movements for even hundreds of images is extremely expensive and time-consuming. Instead, we use the BubbleView \rev{interface} by Kim et al.~\cite{bubbleview,bubble} to record human ``attention'' that is correlated with eye fixations. Unlike eye tracking, which requires expensive equipment and a controlled lab study, BubbleView can be used to to collect large datasets with online crowdsourcing.

In BubbleView, a participant is shown a blurry image 
and can click on different parts of the image to reveal small regions, or \emph{bubbles}, of the image at full resolution. Initial experiments by Kim et al.~\cite{bubble} showed a high correlation between eye fixations collected in the lab and crowdsourced BubbleView click data. In this paper, we confirm this relationship. 

Concurrent work in the computer vision community has applied a similar methodology to natural images. SALICON~\cite{jiang2015salicon} is a crowdsourced dataset of mouse movements on natural images that has been shown to approximate free-viewing eye fixations. Current state-of-the-art models on saliency benchmarks have all been trained on the SALICON data~\cite{cornia2016deep,huang2015salicon,kummerer2014deep,pan2016shallow,zhao2015saliency}. BubbleView was concurrently developed~\cite{bubble} 
to approximate eye fixations on data visualizations with a description task. Some advantages of BubbleView over SALICON are discussed in \cite{bubbleview}.
\rev{Using Amazon's Mechanical Turk (MTurk),}
we collected BubbleView data on a set of 1,411 data visualizations from the MASSVIS dataset~\cite{borkin2013mem}, \rev{spanning a diverse collection of sources (news media, government publications, etc.) and encoding types (bar graphs, treemaps, node-link diagrams, etc.)}. We manually filtered out  visualizations containing illegible and non-English text, as well as scientific and technical visualizations containing too little context.
Images were scaled to have a maximum dimension of 600 pixels to a side while maintaining their aspect-ratios to fit inside the \rev{MTurk} task window.
We blurred the visualizations using a Gaussian filter with a radius of 40 pixels and used
a bubble size with a radius of 32 pixels as in~\cite{bubbleview}. \rev{MTurk} participants were additionally required to provide descriptions for the visualizations to ensure
that they meaningfully explored each image. 
Each visualization was shown to an average of 15 participants. We aggregated the clicks of all participants on each visualization and blurred the click locations with a Gaussian filter with a radius of 32 pixels, to match the format of the eye movement data. 


We used the MASSVIS eye movement data for testing our importance predictions. Fixation maps were created by aggregating eye fixation locations of an average of 16 participants viewing each visualization for 10 seconds. Fixation locations were Gaussian filtered with a blur radius of 32 pixels.
Fig.~\ref{fig:methodology_comparison}a includes a comparison of the BubbleView click maps to eye fixation maps from the MASSVIS dataset.  


\subsection{Ground truth importance for graphic designs}

We used the Graphic Design Importance (GDI) dataset~\cite{odonovan2014} which comes with importance annotations for 1,078 graphic designs from Flickr. Thirty-five \rev{MTurk} participants were asked to label important regions in a design using binary masks, and their annotations were averaged. Participants were not given any instruction as to the meaning of ``importance.'' To determine how BubbleView clicks relate to explicit importance annotations, we ran the BubbleView study on these graphic designs 
and collected data from an average of 15 participants per design. Fig.~\ref{fig:methodology_comparison}b shows comparisons between the GDI annotations and BubbleView click maps. In both data similar elements and regions of designs emerge as important. 

Each representation has potential advantages. The GDI annotations assign a more uniform importance score to whole elements. This can serve as a soft segmentation to facilitate design applications like retargeting. BubbleView maps may be more appropriate for directly modeling human attention. 





\section{Models for predicting importance}

Given a graphic design or data visualization, our task is to predict the importance of the content at each pixel location. We assume the input design/visualization is a bitmap image. The output importance prediction at each pixel $i$ is $P_i\in[0,1]$, where larger values indicate higher importance. 
We approach this problem using deep learning, which has lead to many recent breakthroughs on a variety of image processing tasks in the computer vision community~\cite{NIPS2012_4824,Razavian}, including the closely related task of saliency modeling.

Similar to some top-performing saliency models for natural images~\cite{huang2015salicon,Kruthiventi15}, our architecture is based on
fully convolutional networks (FCNs)~\cite{long2015fully}. 
FCNs are specified by a directed acyclic graph of linear (e.g., convolution) and nonlinear (e.g., max pool, ReLU) operations over the pixel grid, and a set of parameters for the operations.  The network parameters are optimized over a loss function given a labeled training dataset.  We refer the reader to Long et al.~\cite{long2015fully} for more details. 

We predict real-valued importance using a different training loss function from the original FCN work, which predicted discrete object classes.
Given ground truth importances at each pixel $i$,  $Q_i\in[0,1]$, we optimize the sigmoid cross entropy loss for FCN model parameters $\Theta$ over all pixels $i=1,\dots,N$:
\begin{equation}
    L(\Theta) = -\frac{1}{N}\sum_{i=1}^{N}\left(Q_i \log P_i + (1-Q_i) \log(1-P_i)\right)
\label{eqn:loss}
\end{equation}
where $P_i = \sigma\left(f_i(\Theta)\right)$ is the output prediction of the FCN $f_i(\Theta)$ composed with the sigmoid function $\sigma(x)=\left(1+\exp(-x)\right)^{-1}$. Note that the same loss is used for binary classification, where $Q_i\in\{0,1\}$.  Here, we extend it to real-valued $Q_i\in[0,1]$.
We use a different loss than other saliency models based on neural networks 
that optimize Euclidean~\cite{Kruthiventi15,Pan16}, weighted Euclidean~\cite{cornia2016deep}, or binary classification losses~\cite{kummerer2014deep,zhao2015saliency}. Our loss is better suited to $[0,1]$ values, and is equivalent to optimizing the KL loss commonly used for saliency evaluation.

We trained separate networks for data visualizations and for graphic designs. For the data visualizations, we split the 1.4K MASSVIS images for which we collected BubbleView click data into 1,209 training images and 202 test images. For the test set we chose MASSVIS images for which eye movements are available~\cite{beyondMemorability}. For the graphic designs, we split the 1,078 GDI images into 862 training images and 216 test images (80-20\% split). We used the GDI annotations~\cite{odonovan2014} for training. We found that training on the GDI annotations rather than the BubbleView clicks on graphic designs facilitated the design applications better, since the GDI annotations were better aligned to element boundaries.


\begin{figure}
\centering
\includegraphics[width=1\linewidth]{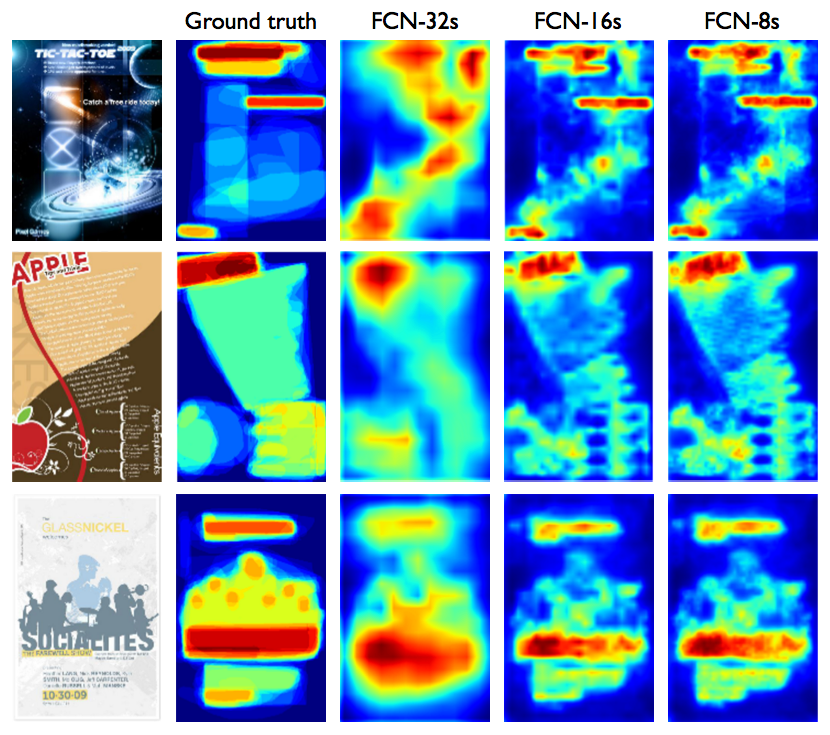}
\caption{We increase the precision of our FCN-32s predictions by combining output from the final layer of the network with outputs from lower levels. The resulting predictions, FCN-16s and FCN-8s, capture finer details.
We found FCN-16s sufficient for our model for graphic designs, as FCN-8s did not add a performance boost. For our model for data visualizations, we found no performance gains beyond FCN-32s.
}
\label{fig:FCNexamples}
\end{figure}

\textbf{Model details:} 
We converted an Oxford VGG-16 convolutional neural network~\cite{simonyan2014very} to an FCN-32s model via network surgery using the 
implementation in Caffe~\cite{jia2014caffe}. The model's predictions are 1/32 of the input image resolution, due to successive pooling layers. To increase the resolution of the predictions and capture fine details, we followed the procedure in Long et al.~\cite{long2015fully} to add skip connections from earlier layers to form FCN-16s and FCN-8s models, that are respectively, 1/16 and 1/8 of the input image resolution. 
We found that the FCN-16s (with a single skip connection from \emph{pool4}) improved the graphic design importance maps relative to the FCN-32s model (Fig.~\ref{fig:FCNexamples}), but that adding an additional skip connection from \emph{pool3} (FCN-8s) performed similarly.  We found that skip connections lead to no gains for the data visualization importance. For our experiments we used the trained FCN-16s for graphic designs and the FCN-32s for data visualizations. 

Since we have limited training data we initialized the network parameters with the pre-trained FCN32s model for semantic segmentation in natural images~\cite{long2015fully}, and fine-tuned it for our task. The convolutional layers at the end of the network and the skip connections were randomly initialized. Training details are provided in the Supplemental Material. 

We opted for a smaller architecture with fewer parameters than some other neural network saliency models for natural images. This makes our model more effective for our datasets, which are currently an order-of-magnitude smaller than the natural image saliency datasets.

\section{Evaluation of model predictions}
\label{sec:evaluation}

We compare the performance of our two importance models to ground truth importance on each dataset. For data visualizations, we compare predicted importance maps to bubble clicks gathered using BubbleView, and to eye fixations from the MASSVIS dataset. For graphic designs, we compare predicted importance maps to GDI annotations.



\subsection{Evaluation criteria}

We evaluate the similarity of our predicted and ground truth importance maps using two metrics commonly used for saliency evaluation~\cite{salMetrics_Bylinskii}: Kullback-Leibler divergence (KL) and cross correlation (CC). CC measures how correlated the pixel-wise values are in the two maps, and treats both false positives and false negatives equally. 
KL, however, measures how well one distribution predicts another. Our importance maps can be interpreted as providing, for each pixel, the probability that the pixel would be considered important by ground truth observers. KL highly penalizes missed predictions, so a sparse map that fails to predict a ground truth important location will receive a large KL value (poor score).
Given the ground truth importance map $Q$ and the predicted importance map $P$, KL is computed as:
\begin{equation}
KL(P,Q) = \sum_{i=1}^{N}\left(Q_i \log Q_i - Q_i \log P_i\right) = L(P,Q) - H(Q),
\end{equation}
%
where $H(Q) = -\sum_{i=1}^{N}\left(Q_i \log Q_i\right)$ is the entropy of the ground truth importance map and $L(P,Q)$ is the cross entropy of the prediction and ground truth. Note the similarity to the loss in Equation~(\ref{eqn:loss}), which is over a Bernoulli random variable; here the random variable is instantiated. A large KL divergence indicates a high dissimilarity between maps, whereas $KL(P,Q)=0$ indicates two maps are identical. KL is in principle unbounded, so to provide a feasible range, we include chance baselines in our experiments.
CC is computed as:
\begin{equation}
CC(P,Q) = \frac{\frac{1}{N}\sum_{i=1}^{N}\left(P_i-\bar{P}\right)\left(Q_i-\bar{Q}\right)}{\sqrt{\frac{1}{N}\sum_{i=1}^{N}\left(P_i-\bar{P}\right)^2}\sqrt{\frac{1}{N}\sum_{i=1}^{N}\left(Q_i-\bar{Q}\right)^2}},
\end{equation}
where $\bar{P} = \frac{1}{N}\sum_{i=1}^{N}P_i$, and respectively for $Q$. CC ranges from -1 to 1, where 1 indicates maximal correlation between two maps $P$ and $Q$. For further intuition about how KL and CC metrics score similarity, \rev{we provide scores above each image} in Fig.~\ref{fig:bestworst_vis}\rev{, and additional examples in the Supplemental Material, showing high- and low-scoring predictions}. 

\subsection{Prediction performance on data visualizations}

\begin{figure}
\centering
\includegraphics[width=1\linewidth]{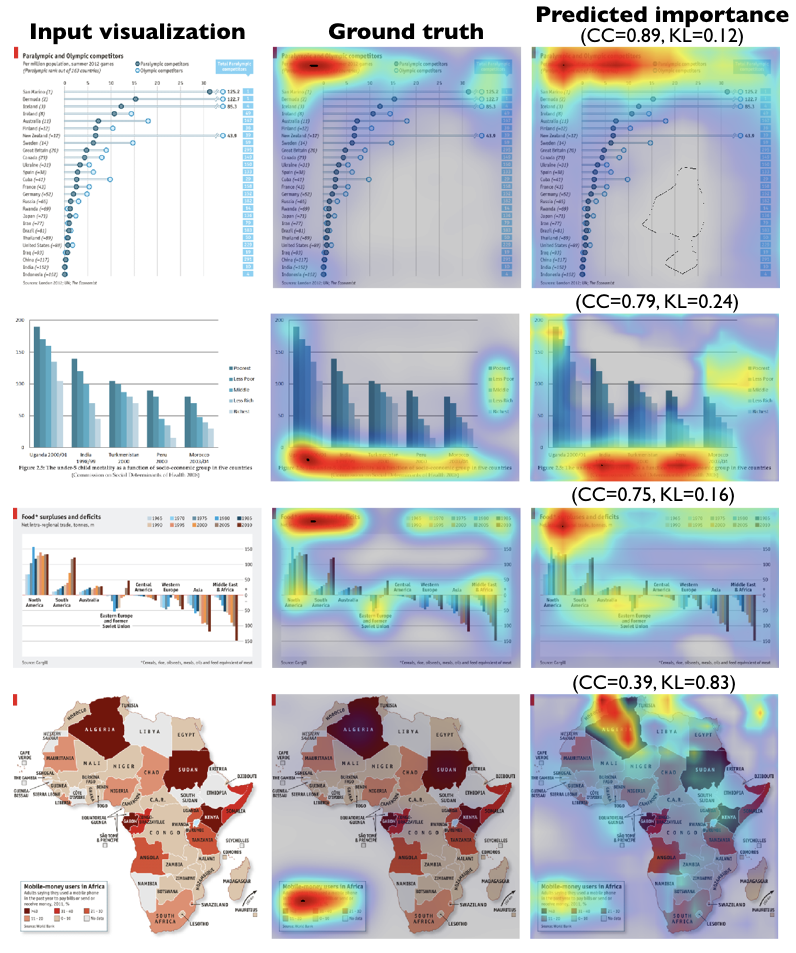}
\caption{Importance predictions for data visualizations, compared to ground truth BubbleView clicks and sorted by performance. Our model is good at localizing visualization titles (the element clicked on, and gazed at, most by human participants) as well as picking up the extreme points on graphs (e.g., top and bottom entries). 
We include a failure case where our model overestimates the importance of the visual map regions. More examples in the Supplemental Material.}
\label{fig:bestworst_vis}
\end{figure}

We include predictions from our importance model in Fig.~\ref{fig:bestworst_vis}.  Notice how we correctly predict important regions in the ground truth corresponding to titles, captions, and legends.  We quantitatively evaluate our approach on our collected dataset of BubbleView clicks.  We report CC and KL scores averaged over our dataset of 202 test images in Table~\ref{tab:datavis_res1}. 

We compare against the following baselines: chance, Judd saliency~\cite{judd2009learning}, and DeepGaze~\cite{kummerer2014deep}, a top neural network saliency model trained on the SALICON dataset~\cite{jiang2015salicon} of mouse movements on natural images. The chance baseline, used in saliency benchmarks~\cite{salMetrics_Bylinskii,Judd_2012}, is computed by uniformly sampling a real value between 0 and 1 at each image pixel. Our approach out-performs all baselines. \rev{KL is highly sensitive to false negatives and drastically penalizes sparser models~\cite{salMetrics_Bylinskii}\footnote{Because of the sensitivity of KL to output regularization, we advise against using it (solely) to compare models~\cite{salMetrics_Bylinskii}.}, explaining the high KL values for DeepGaze in Table~\ref{tab:datavis_res1}. Post-processing or directly optimizing models for specific metrics can yield more favorable performances~\cite{kummerer2017saliency}.}  

\begin{table}
\centering
\begin{tabu} to \textwidth {c c c}
\toprule
Model & CC score $\uparrow$ & KL score $\downarrow$ \\ \midrule
Chance                                  & 0.00     & 0.75     \\
Judd~\cite{judd2009learning}     & 0.11     & 0.49     \\
DeepGaze~\cite{kummerer2014deep} & 0.57     & 3.48     \\
\textbf{Our model}                    & \textbf{0.69}    & \textbf{0.33}     \\ \bottomrule
\end{tabu}
\caption{How well can our importance model predict the BubbleView click maps? We add comparisons to two other top-performing saliency models and a chance baseline. Scores are averaged over 202 test data visualizations. A higher CC score and lower KL score are better.}
 \label{tab:datavis_res1}
\end{table}


How well does our neural network model, trained on clicks, predict eye fixations? 
We find that the predicted importance is representative of eye fixation patterns as well (Table~\ref{tab:datavis_res2})\rev{, although the difference in scores indicates that our model might be learning from patterns in the click data that are different from fixations}.

\begin{table}[]
\centering
\begin{tabu}to \textwidth {c c c}
\toprule
Model              & CC score $\uparrow$     & KL score $\downarrow$      \\ \midrule
Chance             & 0.00          & 1.08          \\
Judd~\cite{judd2009learning}              & 0.19          & 0.74          \\
DeepGaze~\cite{kummerer2014deep}           & 0.53          & 3.10          \\
\textbf{Our model} & \textbf{0.54} & \textbf{0.63} \\
Bubble clicks      & 0.79          & 0.28          \\ \bottomrule
\end{tabu}
\caption{How well can human eye fixations be predicted?
We measured the similarity between human fixation maps and various predictors. Scores are averaged over 202 test data visualizations. A good model achieves a high CC score and low KL score. Our neural \rev{network} model was trained on BubbleView click data, so that is the modality it can predict best. Nevertheless, its predictions are also representative of eye fixation data. As an upper bound on this prediction performance, we consider how well the BubbleView click data predicts eye fixations, and as a lower bound, how well chance predicts eye fixations.}
\label{tab:datavis_res2}
\end{table}

\begin{figure}
\centering
\includegraphics[width=1\linewidth]{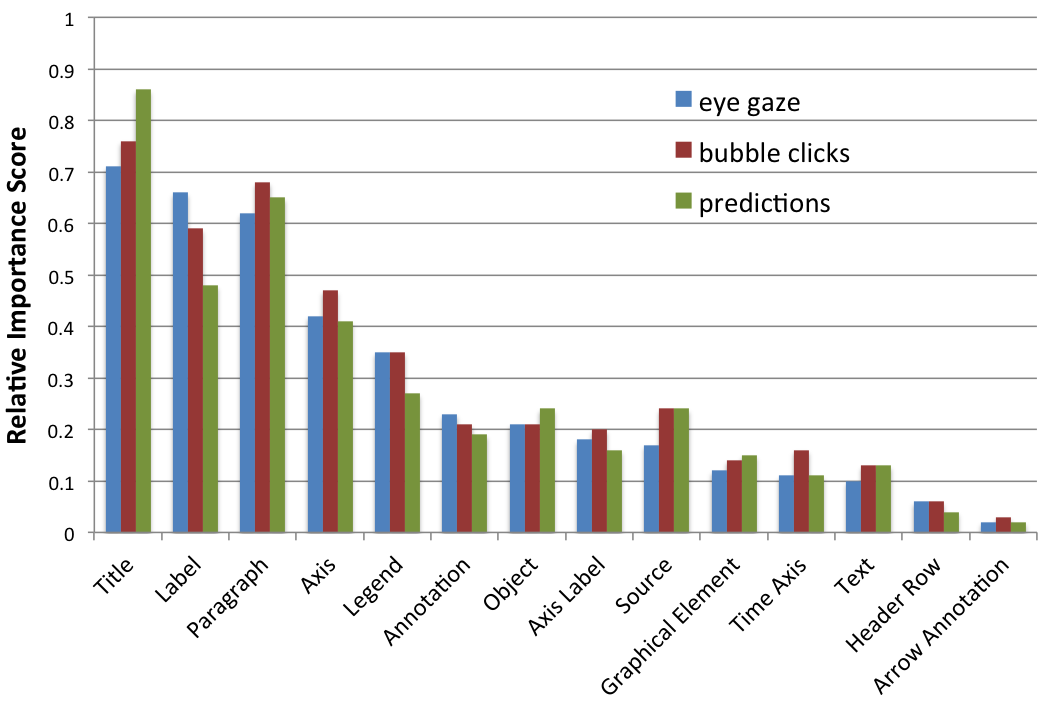}
\caption{Relative importance scores of different elements in a data visualization \rev{assigned by} eye fixation maps, BubbleView click maps, and model predictions.
\rev{Scores were computed by overlapping element segmentations with normalized importance maps, and taking the max of the map within each element as its score.}
The elements that received the most clicks also tended to be highly fixated during viewing (Spearman's $r_s=.96, p<.001$). Text (titles, labels, paragraphs) received a lot of attention. 
Our neural network model correctly predicted the relative importance of these regions relative to eye movements ($r_s=.96, p<.001$).}
\label{fig:viselements}
\end{figure}

\textbf{Which elements are most important?} 
For our analysis, we used the element segmentations available for the visualizations in the MASSVIS dataset~\cite{beyondMemorability}. We overlapped these segmentations with normalized maps of eye fixations, clicks, and predicted importance. We computed the max score of the map within each element to get an importance ranking across elements\footnote{A similar analysis was used to rank the relative importance of objects in natural images~\cite{zoyaECCV2016,jiang2015salicon}.}. 
Text elements, such as titles and captions, were the most looked at\footnote{Among the text and other content in a visualization, titles tend to be best remembered by human observers~\cite{beyondMemorability}.}, and clicked on, elements, and were also predicted most important by our model (Fig.~\ref{fig:viselements}).
Even though our model was trained on BubbleView clicks, the predicted importance remains representative of eye fixation patterns. With regards to differences, our model overpredicts the importance of titles. Our model learns to localize visualization titles very well (Fig.~\ref{fig:bestworst_vis}). 

\subsection{Prediction performance on graphic designs}

The closest approach to ours is the work of O'Donovan et al.~\cite{odonovan2014} who computed an importance model for the GDI dataset. We re-ran their baseline models on the train-test split used for our model (Table~\ref{tab:odonovan_compare}). To replicate their evaluation, we report root-mean-square error (RMSE) and the $R^2$ coefficient, where $R^2=1$ indicates a perfect predictor, and $R^2=0$ is the baseline of predicting the mean importance value (details in Supplemental Material).
The full O'Donovan model (\emph{OD-Full}) \rev{requires manual annotations of} \emph{text}, \emph{face}, and \emph{person} regions, \rev{and would not be practical in an automatic setting}. For a fair comparison, we evaluate our automatic predicted importance model (\emph{Ours}) against the automatic portion of the O'Donovan model, which does not rely on human annotations (\emph{OD-Automatic}). We find that our model outperforms \emph{OD-Automatic}. Our model is \rev{also 100X faster,} since it requires a single feed-forward pass through the network ($\sim$0.1 s/image on a GPU). O'Donovan's method requires separate computations of multiple CPU-based saliency models and image features \rev{($\sim$10 s/image at the most efficient setting)}.

\rev{In Table~\ref{tab:odonovan_compare}, we include the performance of \emph{Ours+OD}, where we added our importance predictions as an additional feature during training of the O'Donovan model, and re-estimated the optimal weights for combining all the features. \emph{Ours+OD} improves upon \emph{OD-Full} indicating that our importance predictions are not fully explainable by the existing features (e.g.,~text or natural image saliency). This full model is included for demonstration purposes only, and is not practical for interactive applications.} 

\begin{figure}
\centering
\includegraphics[width=1\linewidth]{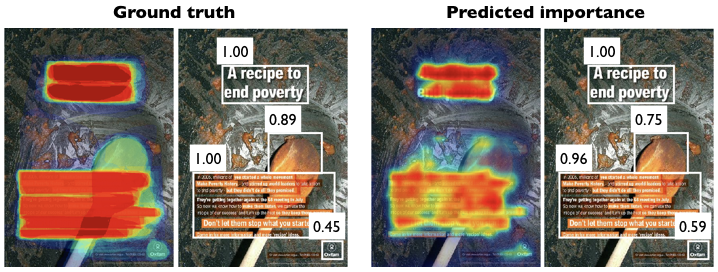}
\caption{An example comparison between the predicted importance of design elements and the ground truth GDI annotations. The heatmaps are overlapped with element bounding boxes and the maximum score per box is used as the element's importance score (between 0 and 1).}
\label{fig:gdibestworst}
\end{figure}

We also annotated elements in each of the test graphic designs using bounding boxes, and computed the maximum importance value in each bounding box as the element's score (Fig.~\ref{fig:gdibestworst}). We obtain an average Spearman rank correlation of 0.56 between the predicted and ground truth scores assigned to the graphic design elements.

\begin{figure}[t!]
\centering
\includegraphics[width=1\linewidth]{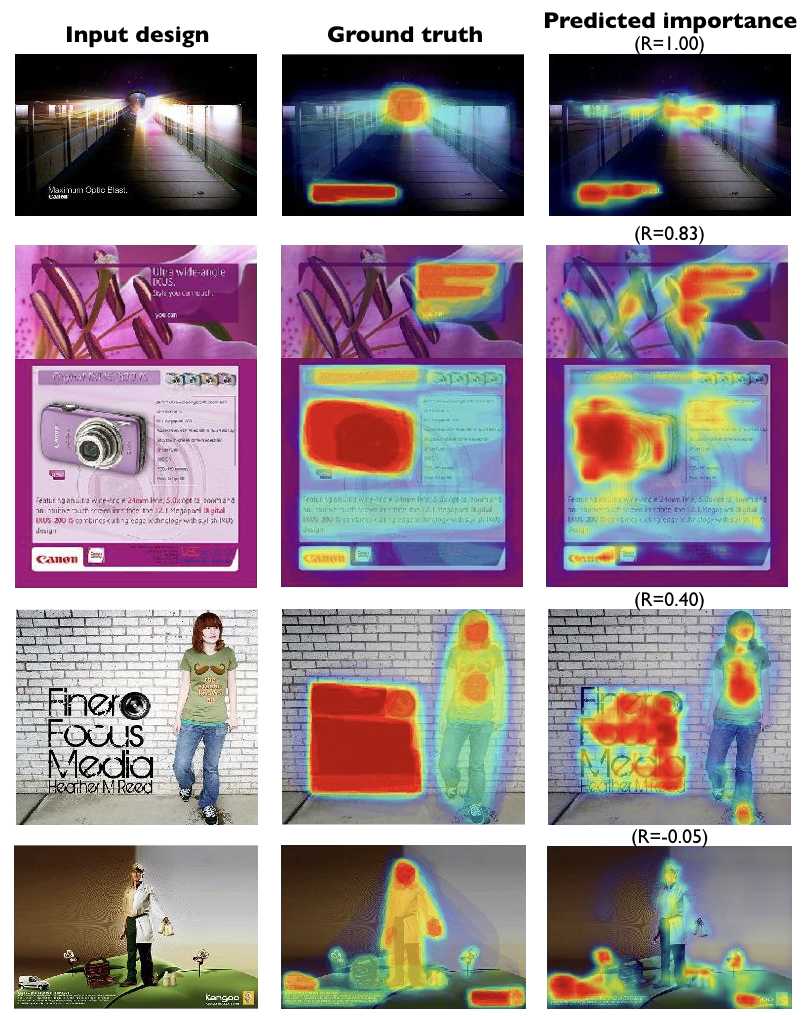}
\caption{Importance predictions for graphic designs, sorted by performance. Performance is measured as the Spearman rank correlation (R) between the importance scores assigned to design elements by ground truth (GDI annotations) and predicted importance maps. A score of 1 indicates a perfect rank correlation; a negative score indicates the element rankings are reversed. The predicted importance maps distribute importance between text and visual features. We include a failure case where the importance of the man in the design is underestimated. More examples in the Supplemental Material.}
\label{fig:gdibestworst2}
\end{figure}

Some examples of predictions are included in Fig.~\ref{fig:gdibestworst2}.
Our predictions capture important general trends, such as larger and more central text and visual elements being more important. However, text regions are not always well segmented (predicted importance is not uniform over a text element), and text written in unusual fonts is not always detected. Such problems could be ameliorated through training on larger datasets. Harder cases are directly comparing the importance of a visual and text, which can depend on the semantics of the text itself (how informative it is) and the quality of the visual (how unexpected, aesthetic, etc.). 


\begin{table}[]
\centering

\begin{tabu} to \linewidth {c c c}
\toprule
Model            & RMSE $\downarrow$          & $R^{2}$ $\uparrow$   \\ \midrule
Saliency         & .229          & .462          \\
OD-Automatic     & .212          & .539          \\
\textbf{Ours}    & \textbf{.203} & \textbf{.576} \\ \hdashline[2pt/2pt]
OD-Full          & .155          & .754          \\
\textbf{Ours+OD} & \textbf{.150} & \textbf{.769} \\ \bottomrule
\end{tabu}
\caption{A comparison of our predicted importance model (\emph{Ours}) with the model of O'Donovan et al.~\protect\cite{odonovan2014}. Lower $RMSE$ and higher $R^{2}$ are better. Our model outperforms the fully automatic O'Donovan variant (\emph{OD-Automatic}). Another fully automatic model from~\protect\cite{odonovan2014} is \emph{Saliency}, a learned combination of 4 saliency models: Itti\&Koch~\protect\cite{itti98model}, Hou\&Zhang~\protect\cite{hou2007saliency}, Judd et al.~\protect\cite{judd2009learning}, and Goferman et al.~\protect\cite{goferman2012context}. We also report the results of the semi-automatic OD-Full model, which includes manual annotations of \emph{text}, \emph{face}, and \emph{person} regions.  When we combine our approach with OD-Full (\emph{Ours+OD}), we can approve upon the OD model. More comparisons are included in the Supplemental Material.}
\label{tab:odonovan_compare}
\end{table}


\subsection{Prediction performance on fine-grained design variations} 
To check for feasibility of an interactive application we perform a more fine-grained test. We want the importance rankings of elements to be adjusted accurately when the user makes changes to their current design. For example, if the user makes a text box larger, then its importance should not go down in the ranking. Our predicted importance model has not been explicitly trained on systematic design variations, so we test if it can generalize to such a setting.

\begin{figure*}
\centering
\includegraphics[width=0.9\linewidth]{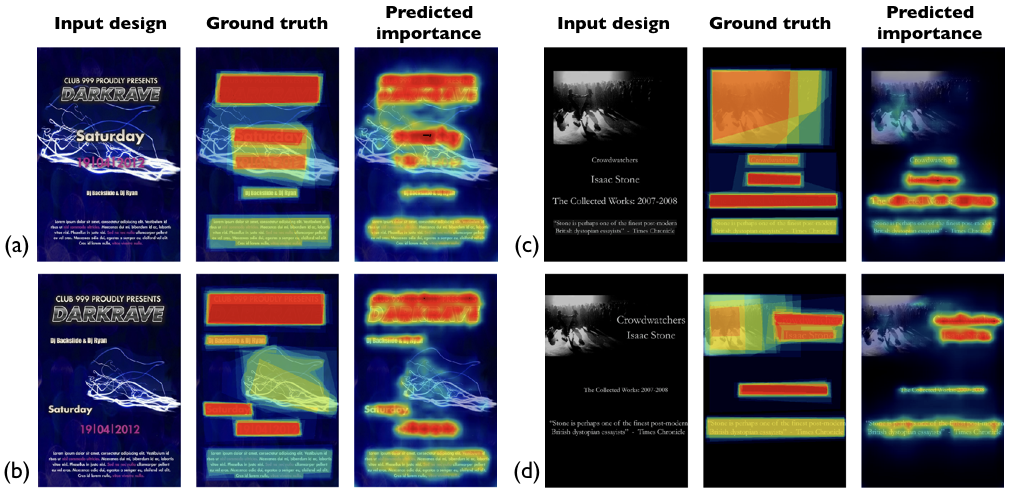}
\caption{Sample input designs, and how the relative importance of the different design elements changes as they are moved, resized, and otherwise modified. For instance, compared to in (a), the event date stands out more and gains importance when it occurs at the bottom of the poster, in large font, on a contrasting background (b). Similarly, when the most important text of the design in (c) is moved to the upper righthand corner where it is not surrounded by other text, it gains prominence (d). Our automatic model makes similar predictions of the relative importance of design elements as ground truth human annotations.}
\label{fig:interactiveeval}
\end{figure*}

We used the Design Improvement Results dataset~\cite{odonovan2014} containing 11 designs with an average of 35 variants. Across variants, the elements are preserved but the location and scale of the elements varies. We repeated the \rev{MTurk} importance labeling task of O'Donovan et al.~\cite{odonovan2014} on a subset of 264 design variants, recruiting an average of 19 participants to annotate the most important regions on each design. 
We averaged all participant annotations per design to obtain ground truth importance heatmaps.
We segmented each design into elements and used the ground truth and predicted importance heatmaps to assign importance scores to all the elements, calculating the maximum heatmap value falling within each segment. The predicted and ground truth importance scores assigned to these elements achieved an average Spearman's correlation $r_s=.53$. As Fig.~\ref{fig:interactiveeval} shows, even though we make some absolute errors, we successfully account for the impact of design changes such as the location and size of various elements. 

\section{Applications}
\label{sec:applications}

We now demonstrate how automatic importance prediction can enable diverse applications. An importance map can provide a common building block for different summarization and retrieval tasks, including retargeting, thumbnailing, and interactive design tools. 
These prototypes are meant as proofs-of-concept, showing that our importance prediction alone can give good results with almost no additional post-processing. 



\subsection{Retargeting}
\label{retargeting}

The retargeting task is to take a graphic design as input, and to produce a new version of that design with specific dimensions. 
Retargeting is a common task for modern designers, who must work with many different output dimensions. 
There is a substantial amount of work on automatic retargeting for natural images, e.g.,  \cite{AS07,Rubinstein10Comparative}. Several of these methods have shown that saliency or gaze provide good cues for retargeting, to avoid cropping out image content that people are likely to pay most attention to, such as faces in photographs. 

The only previous work on retargeting graphic designs is by O'Donovan et al.~\cite{odonovan2014}. They assumed knowledge of the underlying vector representation of the design and used an expensive optimization with many different energy terms. The method we propose uses bitmap data as input, and is much simpler, without requiring any manual annotations of the input image. 

Importance-based retargeting for graphic designs should preserve the most important regions of a design, such as the title and key visual elements. Given a graphic design bitmap as input and specific dimensions, we use the predicted importance map to automatically select a crop of the image with highest importance (Fig.~\ref{fig:retargeting}). Alternative variants of retargeting (e.g., seam carving) are discussed in the Supplemental Material.

\begin{figure*}
\centering
\includegraphics[width=0.9\linewidth]{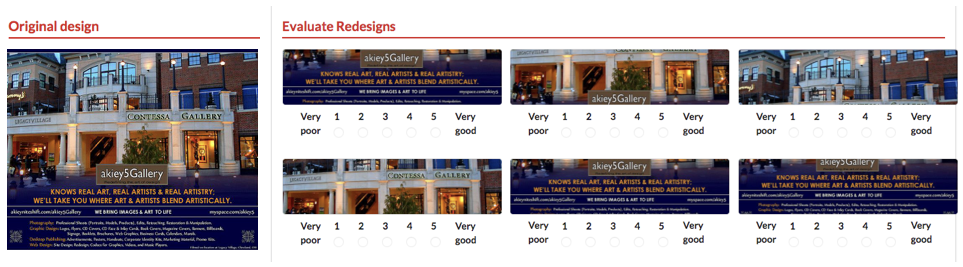}
\caption{\rev{MTurk} interface for evaluating retargeting results of predicted importance compared to other baselines. More experimental details are provided in the Supplemental Material.}
\label{fig:mturk_retargeting}
\end{figure*}

\textbf{Evaluation:} We ran \rev{MTurk} experiments where 96 participants \rev{were presented with a design and 6 retargeted variants}, and were asked to score each variant using a 5-point Likert scale with 1 = very poor and 5 = very good (Fig.~\ref{fig:mturk_retargeting}).
\rev{Each participant completed this task for 12 designs: 10 randomly selected from a collection of 216 designs, and another 2 designs used for quality control.}
We used this task to compare crops retargeted using predicted importance to crops retargeted using ground truth GDI annotations, Judd saliency~\cite{judd2009learning}, DeepGaze saliency~\cite{kummerer2014deep}, and an edge energy map. \rev{We extracted a crop with an aspect ratio of 1:4 from a design using the highest-valued region, as assigned by each of the saliency/importance maps.}
As a baseline, we selected a random crop location. 

\rev{After an analysis of variance showed a significant effect of retargeting method on score, we performed Bonferonni paired t-tests on the scores of different methods.} \rev{Across all 216 designs}, crops obtained using ground truth GDI annotations had the highest score (Mean: 3.19), followed by DeepGaze (Mean: 2.95) and predicted importance (Mean: 2.92). However, the difference between the latter pair of models was not statistically significant. Edge energy maps (Mean: 2.66) were worse\rev{, but not significantly}; while Judd saliency (Mean: 2.47) and the random crop baseline (Mean: 2.23) were \rev{significantly worse in pairwise comparisons with} all the other methods \rev{($p<.01$ for all pairs)}. Results of additional experimental variants are reported in the Supplemental Material.

Our predicted importance outperforms Judd saliency, a natural saliency model commonly used for comparison~\cite{sharonLin,odonovan2014}. Judd saliency has no notion of text. Predicted importance, trained on less than 1K graphic design images, performs on par with DeepGaze, the currently top-performing neural network-based saliency model~\cite{mit-saliency-benchmark} which has been trained on 10K natural images, including images with text. Both significantly outperform the edge energy map, which is a common baseline for retargeting. These results show the potential use case of predicted importance for a retargeting task, even without any post-processing steps.

\subsection{Thumbnailing}
\label{ssec:thumbnailing}

Thumbnailing is similar to retargeting, but with a different goal. It aims to provide a visual summary for an image to make it easier to find relevant images in a large collection \cite{jiao2010visual,teevan2009visual}. 
Unlike previous methods, our approach operates directly on a bitmap input, rather than requiring a specialized representation as input. For this example our domain is data visualizations rather than graphic designs.

Given a data visualization and an automatically-computed importance map as input, we generate a thumbnail by carving out the less important regions of the image. The importance map is used as an energy function, whereby we iteratively remove image regions with least energy first. 
Rows and columns of pixels are removed until the desired proportions are achieved, in this case a square thumbnail. This is similar to seam carving~\cite{AS07,Rubinstein10Comparative}, but using straight seams, found to work better in our setting.
The boundaries of the remaining elements are blurred using the importance map as an alpha-mask with a fade to white.
Qualitatively, the resulting thumbnails consist of titles and other main supporting text, as well as data extremes (from the top and bottom of a table, for instance, or from the left and right sides of a plot). 



\begin{figure*}
\centering
\includegraphics[width=0.7\linewidth]{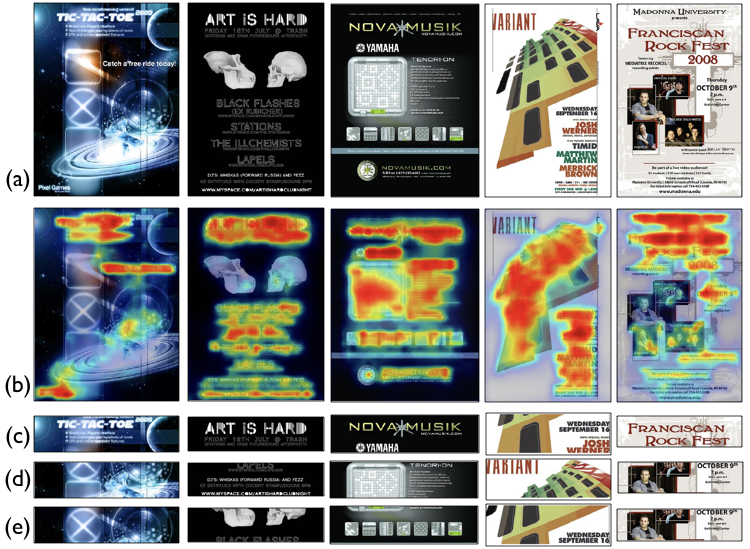}
\caption{(a) Input designs, (b) our predicted importance maps, and (c) automatic retargeting results using the predicted importance maps to crop out design regions with highest overall importance. This is compared to: (d) edge-based retargeting, where gradient magnitudes are used as the energy map, and (e) Judd saliency, a commonly-used natural image saliency model. Additional comparisons are provided in the Supplemental Material.}
\label{fig:retargeting}
\end{figure*}

\begin{figure*}
\centering
(a)\includegraphics[height=5.5cm]{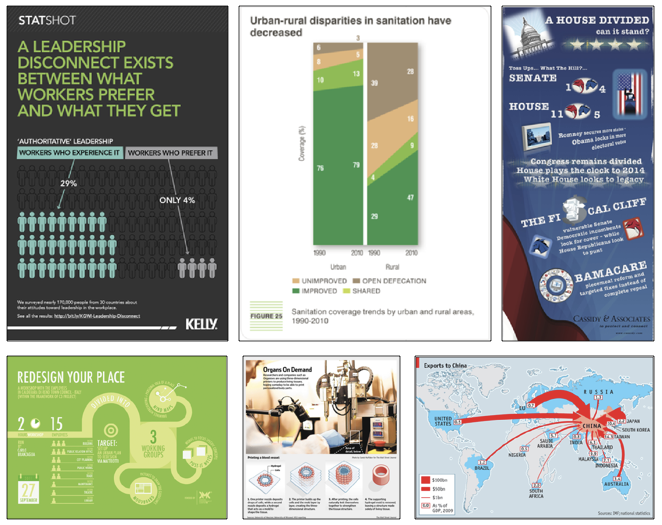}
(b)\includegraphics[height=5.5cm]{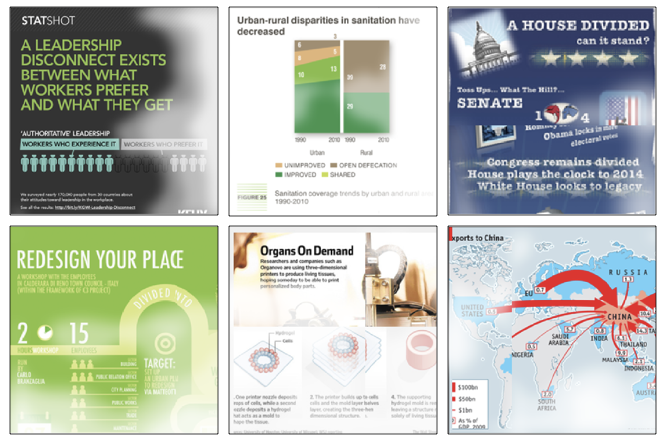}
\caption{Given a set of data visualizations (a), we use our importance maps to automatically generate thumbnails (b). The thumbnails  facilitate visual search through a database of visualizations by summarizing the most important content. More examples can be found in the Supplemental Material.}
\label{fig:thumbnailing}
\end{figure*}

\textbf{Evaluation:} We designed a task intended to imitate a search through a database of visualizations. \rev{MTurk} participants were given a description and a grid of 60 thumbnails, and were instructed to find the visualization that best matches the description. We ran two versions of the study: with the original visualizations resized to thumbnails (Fig.~\ref{fig:thumbnailing}a), and another with our automatically-computed importance-based thumbnails (Fig.~\ref{fig:thumbnailing}b). 
We measured how many clicks it took for participants to find the visualization corresponding to the description in each version. 

\rev{
A total of 400 participants were recruited for our study. After filtering, we compared the performance of 200 participants who performed the study with resized visualizations and 169 participants who saw the importance-based thumbnails.
}

Each \rev{MTurk} assignment, containing a single search task assigned to a single participant, was treated as a repeated observation. We ran an unpaired two-sample t-test to compare the task performance of both groups.
\rev{On average, participants found the visualization corresponding to the description in fewer clicks using the importance-based thumbnails (1.96 clicks) versus using the resized visualizations (3.25 clicks, t(367) = 5.10, p < .001).}
\rev{Our importance-based thumbnails facilitated speedier retrieval, indicating that the thumbnails captured visualization content relevant for retrieval.}

\subsection{Interactive applications}
\label{ssec:interactiveapps}

An attractive aspect of neural network models is their fast run-time performance (Table~\ref{tab:timing}). 
As a prototype, we integrated our importance prediction with a simple design layout tool that allows users to move and resize elements, as well as change color, text font, and opacity (Fig.~\ref{fig:interactiveapp}). With each change in the design, an importance map is recomputed automatically to provide immediate feedback to the user. 
The accompanying video \rev{and demo (\url{visimportance.csail.mit.edu})} demonstrate the interactive capabilities of our predictions.  Our experiments in the \emph{Evaluation} section provide initial evidence that our model can generalize to the kind of fine-grained design manipulations, like the resizing and relocation of design elements, that would be common in an interactive setting. 
Determining how best to use importance prediction to provide feedback to users is an interesting problem for future work. For example, importance prediction could help in formulating automatic suggestions for novice users to improve their designs.

\rev{
\textbf{Timing:} On a Titan-X GPU, our model computes the importance map for a design in the GDI dataset (600$\times$450 pixels) in 100 ms. Table~\ref{tab:timing} provides some timing information for our model on differently-sized images.
}


\begin{table}[h!]
\centering
\begin{tabu} to \textwidth {c c c c c c}
\toprule
Image size (pixels/side) & 300 & 600 & 900 & 1200 & 1500 \\ \midrule
Avg. compute time (ms)         & 46  & 118 & 219 & 367  & 562  \\ \bottomrule
\end{tabu}%
\caption{\rev{Time (in milliseconds) taken by our model to compute an importance map for differently-sized images, averaged over 100 trials.}}
\label{tab:timing}
\end{table}













\rev{
\section{Limitations}
Our neural network model is only as good as the training data we provide it. In the case of data visualizations, there is a strong bias, both by the model and the ground truth human data, to focus on the text regions. This behavior might not generalize to other types of visualizations and tasks. Click data, gathered via the BubbleView interface, is not uniform over elements, unlike explicit bounding box annotations (i.e., as in the GDI dataset~\cite{odonovan2014}). While this might be a better approximation to natural viewing, non-uniform importance across design elements might cause side-effects for downstream applications like thumbnailing, by cutting off parts of elements or text. 
}

\section{Conclusions}

We curated hundreds of examples of graphic designs and data visualizations, annotated with importance, to train fully convolutional neural network models to predict importance maps for novel designs. We showed that our computational predictions approximate ground truth human data enough to be used for a number of automatic applications. Our importance maps act as a common underlying representation for retargeting of graphic designs, thumbnailing 
of data visualizations, and in a prototype interactive design application. 


This paper presents the first neural network model for predicting saliency or importance in graphic designs and data visualizations, capable of generalizing to a wide range of design formats.
Moreover, the fast test-time performance of our model makes it feasible for the predictions to be used in interactive design tools. 
Our approach is not limited to graphic designs and data visualizations. The methodology 
and models can easily be adapted to other visual domains, such as websites~\cite{bubbleview}. As better webcam-based eyetracking methods become available (e.g., \cite{krafka2016eye,papoutsakiwebgazer,xu2015turkergaze}) possibilities also open up for directly training our model from eye movement data. Future work can also explore the use of importance predictions to offer more targeted design feedback and to provide automated suggestions to a user.

\section{Acknowledgements}
We would like to thank Joel Brandt and the anonymous reviewers for useful feedback. Thank you to Matthias K\"{u}mmerer and Adri\`{a} Recasens for help computing saliency models. ZB would like the acknowledge the support of the Natural Sciences and Engineering Research Council of Canada
Postgraduate Doctoral Scholarship (NSERC PGS-D). 

\bibliographystyle{SIGCHI-Reference-Format}
\bibliography{main.bbl}

\end{document}